\documentclass[]{pasj01}
\draft

\usepackage{mathpazo}
\usepackage[T1]{fontenc} 

\graphicspath{{./eps/}}
\usepackage{siunitx}
\usepackage{color}      

\newcommand{\eV}{\,\si{eV}}
\newcommand{\keV}{\,\si{keV}}
\newcommand{\nm}{\,\si{nm}}
\newcommand{\cm}{\,\si{cm}}
\newcommand{\km}{\,\si{km}}

\newcommand{\g}{\,\si{g}}

\begin{document} 

\title{ 
Suzaku Detection of Solar Wind Charge Exchange Emission from a Variety of Highly-ionized Ions in an Interplanetary Coronal Mass Ejection
 }

\author{Kazunori \textsc{asakura}\altaffilmark{1}%
\email{asakura\_k@ess.sci.osaka-u.ac.jp}}

\author{Hironori \textsc{matsumoto}\altaffilmark{1,2}}
\author{Koki \textsc{okazaki}\altaffilmark{1}}
\author{Tomokage \textsc{yoneyama}\altaffilmark{1}}
\author{Hirofumi \textsc{noda}\altaffilmark{1,2}}
\author{Kiyoshi \textsc{hayashida}\altaffilmark{1,2,3}}
\author{Hiroshi \textsc{tsunemi}\altaffilmark{1}}
\author{Hiroshi \textsc{nakajima}\altaffilmark{4}}
\author{Satoru \textsc{katsuda}\altaffilmark{5}}
\author{Daiki \textsc{ishi}\altaffilmark{6}}
\author{Yuichiro \textsc{ezoe}\altaffilmark{6}}


\altaffiltext{1}{Department of Earth and Space Science, Graduate School of Science, Osaka University, 1-1 Machikaneyama, Toyonaka, Osaka 560-0043, Japan}
\altaffiltext{2}{Project Research Center for Fundamental Sciences, Graduate School of Science, Osaka University, 1-1 Machikaneyama, Toyonaka, Osaka 560-0043, Japan}
\altaffiltext{3}{Japan Aerospace Exploration Agency, Institute of Space and Astronautical Science, 3-1-1 Yoshinodai, Chuo-ku,Sagamihara, Kanagawa 252-5210, Japan}
\altaffiltext{4}{College of Science and Engineering, Kanto Gakuin University, 1-50-1 Mutsuura Higashi, Kanazawa-ku, Yokohama, Kanagawa 236-8501, Japan}
\altaffiltext{5}{Graduate School of Science and Engineering, Saitama University, 255 Shimo-Ohkubo, Sakura, Saitama 338-8570, Japan}
\altaffiltext{6}{Department of Physics, Tokyo Metropolitan University, 1-1 Minami-Osawa, Hachioji, Tokyo 192-0397, Japan}

\KeyWords{X-rays: diffuse background --- solar wind --- solar--terrestrial relations --- Sun: coronal mass ejections (CMEs) --- Sun: flares --- }  

\maketitle

\begin{abstract}
X-ray emission generated through solar-wind charge exchange (SWCX) is known to contaminate X-ray observation data, the amount of which is often significant or even dominant, particularly in the soft X-ray band, 
when the main target is comparatively weak diffuse sources, depending on the space weather during the observation.
In particular, SWCX events caused by interplanetary coronal mass ejections (ICMEs) tend to be spectrally rich and to provide critical information about the metal abundance in the ICME plasma.
We analyzed the SN1006 background data observed with Suzaku on 2005 September 11 shortly after an X6-class solar flare, signatures of which were separately detected together with an associated ICME. 
We found that the data include emission lines from a variety of highly ionized ions generated through SWCX.
The relative abundances of the detected ions were found to be consistent with those in past ICME-driven SWCX events.
Thus, we conclude that this event was ICME-driven.
In addition, we detected a sulfur \emissiontype{XVI} line for the first time as one from the SWCX emission, the fact of which suggests that it is the most spectrally-rich SWCX event ever observed.
We suggest that observations of ICME-driven SWCX events can provide a unique probe to study the population of highly-ionized ions in the plasma, which is difficult to measure in currently-available in-situ observations.

\end{abstract}

\section{Introduction}


Coronal mass ejections (CMEs) are the most energetic eruptive phenomena in our solar system.
The occurrence rate of CMEs depends on the solar activity.
Notably, there is a linear correlation between the occurrence rate and sunspot number (\cite{Webb1994}). 
A CME transfers a large amount of plasma (typically $10^{15}$--$10^{16}\g$) from the low corona to the solar system (\cite{Colaninno2009}). 
The CMEs which are directed toward the Earth are particularly called ``halo CMEs'', which were named after coronagraph images. 
The ejected plasma of the halo CME tends to hit the Earth and is often observed as a disturbance of solar wind (interplanetary CME, called ICME).
Whereas a variety of properties of the CMEs and ICMEs have been identified (reviewed by, e.g., \cite{Chen2011}; \cite{Webb2012}), precise mechanisms of their initiation and energy-release procedure are not fully understood.

Since the first detection of a CME in 1971 (\cite{Tousey1973}), thousands of CMEs have been detected, mainly in coronagraph observations.
In addition to imaging observations including coronagraph ones, recently-developed ultraviolet spectrometers enable us to obtain new aspects of CMEs, providing essential diagnostic information about the plasma components.
Alternatively, information of ICMEs can be obtained in-situ by spacecrafts.
Arrival of an ICME appears as various forms of observable signatures (see the review by \cite{Zurbuchen2006}).
In some cases, ICMEs have enhanced magnetic structures (termed ``magnetic cloud'' in \cite{Burlaga1981}). 
The magnetic clouds have low proton temperatures and plasma beta (the ratio of the plasma pressure to the magnetic pressure), which are also indicators of the arrival of the ICMEs.

ICMEs have also drawn a significant attention in X-ray astronomical observations because propagating plasma causes geomagnetic storms, which affect observations with X-ray satellites.
In addition to the geomagnetic storms, highly-ionized ions in the plasma are the sources of additional X-rays during observations of celestial objects.
These additional X-rays are ascribed to so-called solar wind charge exchange (SWCX), 
\textit{i.e.}, the phenomenon whereby an electron in a neutral atom is transferred to a highly-ionized ion in solar wind during their collisions.
When an electron in an excited state falls back to the ground state, the energy is released as one or more photons in the extreme ultraviolet to soft X-ray range.
This mechanism was first suggested by \citet{Cravens1997} to explain the line emission spectrum discovered in observations of C/Hyakutake 1996 B2 (\cite{Lisse1996}).
 \citet{Freyberg1998} and \citet{Cox1998} suggested that solar wind could produce X-rays in the exosphere of the Earth (geocoronal SWCX) and in the heliosphere (heliospheric SWCX) and 
that the mysterious X-ray time variation detected during the ROSAT All-Sky Survey, which was reported by \citet{Snowden1994} and dubbed as Long-Term Enhancement (LTE), might have originated in the SWCX.
Since solar wind, which continually blows into the heliosphere, always interacts more or less with neutral atoms in the solar system, the SWCX is now recognized as a persistent background component in observations of X-ray diffuse sources. 

To date, plenty of SWCX detections with Chandra, XMM-Newton, and Suzaku have been reported (e.g., \cite{Wargelin2004}; \cite{Snowden2004}; \cite{Fujimoto2007}; for more details see \cite{Kuntz2019}).
Systematic studies of the time variations of the soft X-ray flux have been conducted with large archival datasets of XMM-Newton and Suzaku (\cite{Carter2008}; \cite{Carter2011}; \cite{Ishi2017}).
Roughly a hundred observational datasets in each of the XMM-Newton and Suzaku archives have been found to show time variations caused by the geocoronal SWCX.
This implies that the data that contain time-variable SWCX components are not rare, hence the significance of the study of the SWCX to handle X-ray observational data correctly.
While an increasing number of the samples of the SWCX are available by now, it still remains difficult to construct a comprehensive theoretical model that describes the observed properties of SWCX observations, such as times series and overall flux levels, for arbitrary spacecraft look directions and epochs (\cite{Kuntz2019}).
One of the major difficulties stems from the fact that the properties of ions in solar wind vary with both solar cycle and solar latitude; the intensity and fluctuation of SWCX emission are highly dependent on the viewing point.
However, the global structure of the solar wind, especially at a high solar latitude, cannot be clarified with current in-situ measurements.
Further X-ray observation in a variety of directions is the only way to refine the entire model including the high solar latitude so far.

In addition to the evaluation of the SWCX model, SWCX events associated with ICMEs also provide us with ionic information of the plasma from various emission lines.
The most spectrally-rich case of the SWCX events reported so far was that by \citet{Carter2010}, where they made systematic study, using the XMM-Newton archive, and made unequivocal detections of
 the lines from various ions including highly ionized neon, magnesium, and silicon.
Their work is a good precedent that the components of ICME plasma can be identified indirectly with X-ray observations.
One of the notable characteristics of the event is that the source ICME was accompanied with an X-class flare.
\citet{Reinard2005} suggested a positive correlation between the solar flare magnitude and ionization state of the CME plasma.
 Providing this correlation is correct, the ICMEs associated with large flares should yield spectrally-rich SWCX events, which have plenty of information about the ICME plasma.
However, the opportunities to detect such spectrally-rich events rarely arise because X-class flares are rare in the first place and because not all the X-class flares are associated with CMEs.
Thus, the SWCX events caused by ICMEs associated with X-class flares are valuable samples.

In this paper, we report a newly discovered SWCX emission from the SN1006 background data observed by Suzaku, which is likely to be associated with an ICME accompanied with an X-class flare.
This event could not be observed with XMM-Newton unfortunately, due to the intense solar activity.
Conventionally, SWCX emission can be simply identified as a temporal excess from the stable component in the observed X-ray light-curve.
However, this simple method is not applicable to the dataset like the said data, where no X-ray time variation is observed within the observation period.
Our strategy is to compare the data with another data obtained in the same region at a different observation epoch to subtract the stable component and identify a potential excess.
We summarize the observations and our data analysis flow in \S2 and \S3, respectively.
Discussion about the correlation with a CME and significance of this detection is given in \S4.

\section{Observations}

\subsection{SN1006 background}

\begin{figure}
\begin{center}
\includegraphics[width=18cm]{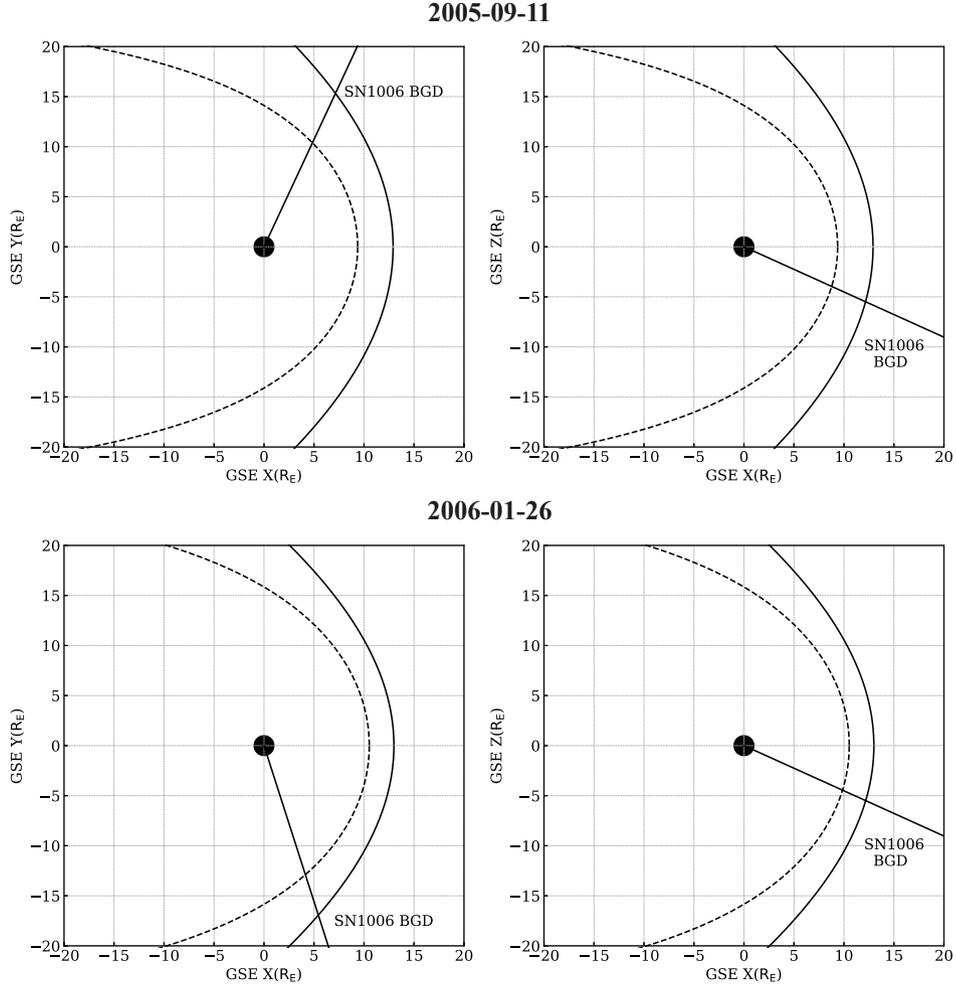}
\end{center}
\caption{Line of sight directions in the GSE X-Y and X-Z planes during each Suzaku observation. Dashed and solid curves show the positions of the magnetopause and the bow shock at the end of each observational period, respectively.}
\label{fig:GSE}
\end{figure}

We used the Suzaku data from the region ``SN1006 background'', located at $(\alpha, \delta)_{\mathrm{J2000.0}} = (224.65^{\circ}, -42.40^{\circ})$.
The primary purpose of the original observations was to take dataset served as the background for the nearby main target SN1006.
This region was observed twice by Suzaku, on 2005 September 11 and 2006 January 26.
The average normalized pointing vectors in the Geocentric Solar Ecliptic (GSE) coordinates were (0.7704, 0.8259, $-$0.4117) and (0.2789, $-$0.8676, $-$0.4117) in the first and second observations, respectively.
Figure~\ref{fig:GSE} shows the spacecraft pointing direction along with calculated models of the magnetopause (\cite{Shue1998}) and the bow shock ( \cite{Merka2005}) for the epochs corresponding to the end of the observational periods.

Suzaku (\cite{Mitsuda2007}), the 5th Japanese X-ray satellite, had the two types of main instruments; the X-ray Imaging Spectrometer (XIS; \cite{Koyama2007}) and the Hard X-ray Detector (HXD; \cite{Takahashi2007}).
Suzaku has four XISs, each of which is an X-ray charge-coupled device (CCD) and works as the focal-plane detector of the X-Ray Telescope (XRT; \cite{Serlrmitsos2007}) for each.
Three of the four XISs are front-illuminated CCDs (FI CCDs; XIS0, XIS2, XIS3), whereas the other is a back-illuminated CCD (BI CCD; XIS1).
These X-ray CCDs have a relatively low and stable background level, owing to the low Earth orbit of $550\km$ in altitude, and thus are suitable to detect SWCX emission.
The observational parameters are summarized in table \ref{table:obs_data}.

\begin{table}
  \tbl{Summary of the observational parameters.}{%
  \begin{tabular}{ccc}
        \hline \hline
      Target name                                       & \multicolumn{2}{c}{SN 1006 SW BG}   \\
      Observation ID                                    &                    100019040                      &                      100019060                    \\
      Observation start (UT)                        &               2005-09-11 23:59                  &                 2006-01-26 17:16               \\
      Effective exposure (ks)                       &                         25.2                            &                           17.0                          \\
      Target coordinates ($\alpha, \delta)$(J2000) & ($224^{\circ}.6550, -42^{\circ}.4005$) & ($224^{\circ}.6468, -42^{\circ}.4025$)\\
      \hline \hline
    \end{tabular}}\label{table:obs_data}
\end{table}

\subsection{Solar activity and geomagnetic field}

\begin{figure}
\begin{center}
\includegraphics[width=17cm]{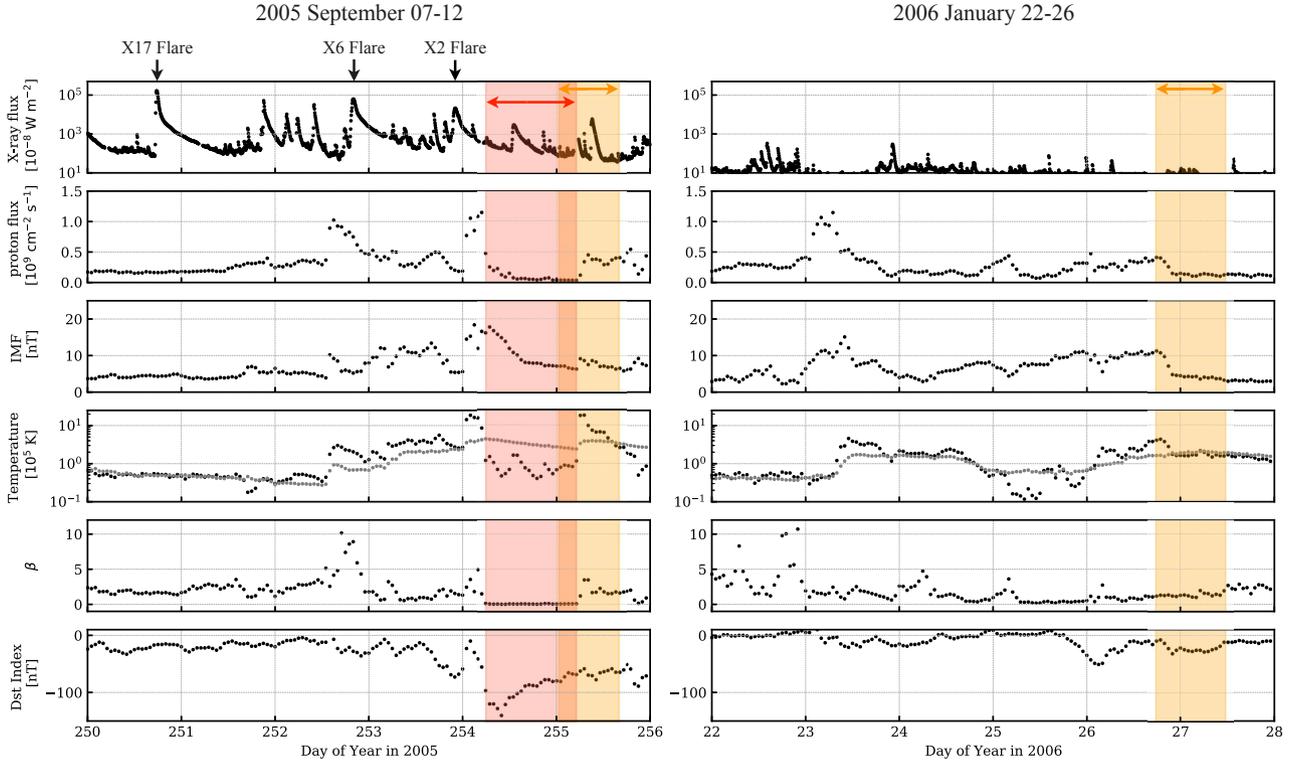}
\end{center}
\caption{From the top panel, solar X-ray light curve ($0.1$--$0.8\nm$), solar proton flux, the IMF, proton temperature, plasma beta, and Dst Index. Gray points in the fourth panel indicate the temperature $T_\mathrm{exp}$ expected from the solar wind velocity, calculated with the empirical formula in \citet{Lopez1987}. The orange areas indicate the periods of the Suzaku observations. The plotted points in the first and other panels are the average values over 5-min and 1-hour periods, respectively. Red shaded region shows the period of a lower plasma beta, starting on DOY 254 in 2005. }
\label{fig:SWobservables}
\end{figure}

Shortly before the first observation (i.e., in early September 2005), several successive X-class solar flares occurred although it was close to solar-cycle minimum.
In particular, a solar flare on 2005 September 7 was huge, classified as the X17 class. 
Subsequently, X6- and X2-class flares erupted on September 9 and 10. 
These X-class flares entailed CMEs, as reported in \citet{Wang2006}.
We plot the time-series of some selected parameters of the solar wind during the Suzaku observations in figure \ref{fig:SWobservables}:
 the solar X-ray flux in the $0.1$--$0.8\nm$ band observed by Geostationary Operational Environmental Satellite (GOES),
 solar proton flux, interplanetary magnetic field (IMF), proton temperature, plasma beta measured by WIND \footnote{We obtained the data from OMNIWeb; https://omniweb.gsfc.nasa.gov},
 and Dst Index provided by the World Data Center for Geomagnetism, Kyoto, Japan. 
Gray points in the fourth panel of figure \ref{fig:SWobservables} indicate the temperature $T_\mathrm{exp}$ expected from the solar wind velocity, calculated with the empirical formula in \citet{Lopez1987}.

At the beginning of day of year (DOY) 254 in 2005, the plasma beta decreased and measured proton temperature dropped below $T_\mathrm{exp}$ for a while, 
which followed discontinuous rises of the proton flux and proton temperature at DOY 254.0 (indicated by the red shaded region in figure \ref{fig:SWobservables}).
This phenomenon of a sudden drop of the plasma beta is a well-known indicator of a front of an ICME passing the observer's location; 
the enhancements of the proton flux and temperature are due to a CME-driven front shock, whereas the low proton temperature and low plasma beta are caused by a magnetic cloud.
This CME event was suspected to be caused by the X6 flare on September 9, which propagated from the sun to the L1 point (where the GOES and WIND satellites stayed) with a velocity of $\sim1400\,\mathrm{km\,s^{-1}}$ (\cite{Wang2006}).
The mass of the CME is $\sim1.6\times10^{17}\g$, which is ranked the most massive halo CME in the SOHO LASCO CME catalog (\cite{Gopalswamy2009}).\footnote{https://cdaw.gsfc.nasa.gov/CME\_list/}
From these facts, we conclude that the Suzaku observation on 2005 September 11 was affected by the CME, whereas that on 2006 January 26 was not. 
Hereafter, we refer to the first and second Suzaku observation periods as the \textit{active} and \textit{stable} periods, respectively.

\section{Analysis}

We analyzed the data obtained with the XISs in both the \textit{active} and \textit{stable} periods.
In our data reduction and analysis, we utilized the software included in the HEAsoft package version 6.25.
We used the 3$\times$3 and 5$\times$5 editing modes for each of the XIS cleaned event data 
(filtered with the standard screening process\footnote{https://heasarc.gsfc.nasa.gov/docs/suzaku/processing/criteria\_xis.html}; processing script version is 3.0.22.43).
Hereafter, quoted errors refer to 90\% uncertainties unless otherwise noted.

\subsection{XIS1 images}

\begin{figure}
 \begin{center}
  \includegraphics[width=17cm]{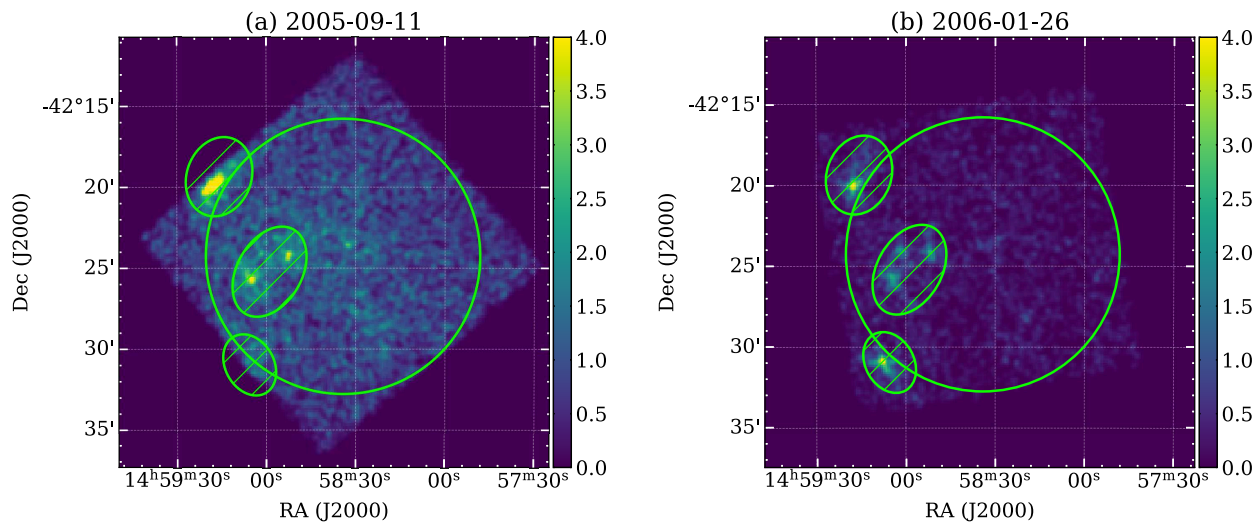}
 \end{center}
\caption{XIS1 $0.4$--$2.0\keV$~images in the (a) \textit{active} and \textit{stable} periods. The color scale represents counts per pixel on a linear scale. The green circles show the region where we extracted the X-ray events (we excluded the hatched regions to get rid of the few point sources).}
\label{fig:xis1_image}
\end{figure}

We examined the data of the BI CCD (XIS1), which is more sensitive than the FI CCDs (XIS0, 2, and 3) in the soft ($<1\keV$) X-ray band, in which SWCX emission was mainly observed.
Figure \ref{fig:xis1_image} shows XIS1 images in the $0.4$--$2.0\keV$~band, without any background subtraction.
The X-ray image of the \textit{active} period was found to be clearly brighter than that of the \textit{stable} period.
Therefore, some additional X-ray emission must exist in the \textit{active} period.
We extracted the X-ray events from a circular region with a radius of 8.5 arcmin, excluding regions encompassing a few point sources (hatched regions in figure~\ref{fig:xis1_image}).  
We confirmed the same X-ray enhancements in the FI CCDs (XIS0, XIS2, and XIS3). Hence, the X-ray enhancement was not instrumental.

\subsection{Light Curve}

\begin{figure}
 \begin{center}
  \includegraphics[width=17cm]{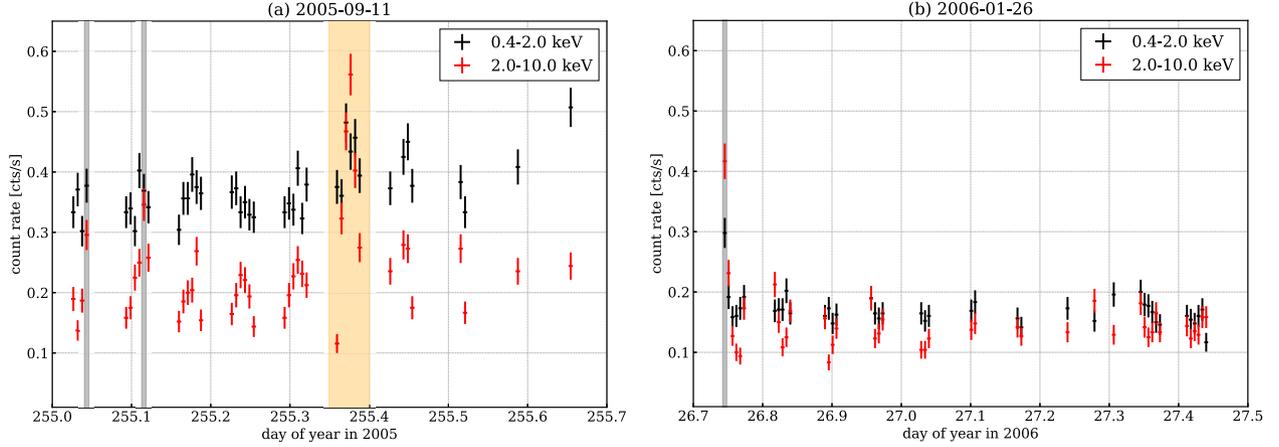}
 \end{center}
\caption{XIS1 480\,s-bin light curves in the (a) \textit{active} and (b) \textit{stable} periods in the (black) $0.4$--$2.0\keV$ and (red) $2.0$--$10.0\keV$~bands. Orange and gray shaded regions show  the ``flare''  and a few rapid-rise periods, respectively, the data from which are excluded in the analyses.}
\label{fig:xis1_lc}
\end{figure}

We extracted 480\,s-bin light curves in the $0.4$--$2.0\keV$ and $2.0$--$10.0\keV$~bands for each XIS individually, and plot those of the XIS1 in figure \ref{fig:xis1_lc}.
The $2.0$--$10.0\keV$ light curves of any of the XISs showed a distinctive X-ray enhancement between DOYs 255.35 and 255.40 (indicated with the orange-shaded area in figure~\ref{fig:xis1_lc}).
The timing coincided with a known M6-class solar flare (we can confirm the flare in the top panel of figure \ref{fig:SWobservables}).
Thus, we excluded the data in this time region in the following analyses. 
The $2.0$--$10.0\keV$ light-curves of the XIS1 and 2 showed also a couple of sharp spikes with a shorter time scale than that of the typical solar flare.
However, the spikes did not appear in the XIS0 or XIS3 light curves. 
The inconsistency among the detectors implied that the spikes originated in something instrumental. 
To identify quantitatively the time regions of the instrumental spikes, we calculated the average and the standard deviation of the count rate in the $2.0$--$10.0\keV$~band, excluding the above-mentioned solar-flare period. 
Then, we selected the time bins in the light curves the count rate of which were out of the 90\% confidence interval as the anomaly bins and excluded them in the following analyses. 
The excluded time bins are indicated with gray-shaded areas in figure \ref{fig:xis1_lc}.
\newline\indent We found that the average count rate of the $0.4$--$2.0\keV$~light curves in the \textit{active} period was apparently nearly twice as high as that in the \textit{stable} period (figure~\ref{fig:xis1_lc}).
We can also confirm the slight increase of the count rate of the $2.0$--$10.0\keV$~light curves, even excluding the solar-flare period (a possible origin of the increase are described in \S3.3).

\subsection{Spectral Fitting}

We extracted X-ray events from each XIS data and made BI  (XIS1) and FI  (XIS0 + XIS2 + XIS3) spectra in each observation period. 
The FI and BI spectra are binned with minima of 50 and 30 counts, respectively.
Then we subtracted from each of the FI and BI spectra a simulated non X-ray background (NXB) spectrum, which was created with \verb|xisnxbgen| (\cite{Tawa2008}).
We made the redistribution matrix files (RMFs), using \verb|xisrmfgen|, and the auxiliary response files (ARFs), using \verb|xissimarfgen| (\cite{Ishisaki2007}) under the assumption
 that the emission region is a circular region with a radius of 20 arcmin, given that the X-ray emission from SN1006 background area appeared uniform.
For spectral fitting in this work, we used \verb|XSPEC| package (version 12.10.1).

\subsubsection{\textit{stable} period}

We analyzed the BI and FI spectra in the \textit{stable} period with model fitting.
In general, the NXB-subtracted X-ray spectrum of blank sky consists mainly of three components: the Galactic Halo (GH), the Local Hot Bubble (LHB), and the cosmic X-ray background (CXB) (e.g., \cite{Kushino2002}).
The GH and LHB are thin thermal plasma (with typical temperatures of $kT \sim 0.3\keV$ and $\sim 0.1\keV$, respectively).
The CXB is now understood as the collective emission from unresolved active galactic nuclei (AGNs) and its emission is well approximated by a power-law with a photon index of 1.41 (\cite{Kushino2002, Luca2004}). 

\begin{figure}
 \begin{center}
  \includegraphics[width=13cm]{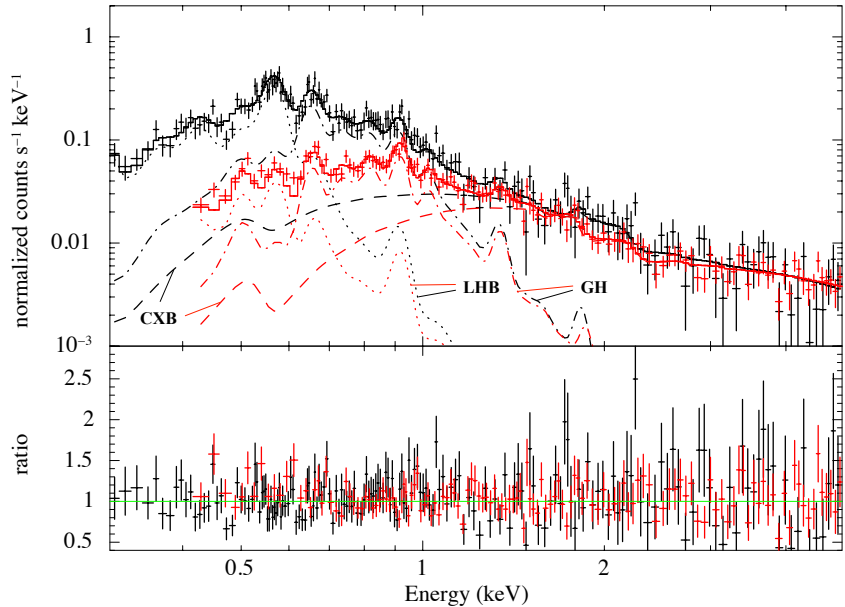}
 \end{center}
\caption{Spectra obtained with the (black) BI-CCD and (red) FI CCDs in the \textit{stable} period. 
Solid lines show the best-fit models of the (dashed-dotted lines) GH, (dotted lines) LHB, and (dashed lines) CXB components.}
\label{fig:stable_spec}
\end{figure}

Considering the composition, we first fitted the $2.0$--$5.0\keV$~band in the spectra, which is likely to be dominated by the CXB, with a power-law with the photon index  fixed to 1.41.  
The normalization of the power-law was determined to be $10.1 \mathrm{\,photons\,s^{-1}\,cm^{-2}\,keV^{-1}\,str^{-1}}$ at $1\keV$, which is consistent with the past result obtained with ASCA (\cite{Kushino2002}).
Then, we extended the energy ranges to $0.3$--$5.0\keV$ and $0.4$--$5.0\keV$~for the BI and FI spectra, respectively, for the further analysis. 
We adopted the two thin-thermal models and a power-law (\verb|phabs*(vapec+powerlaw)+vapec| in \verb|XSPEC|). 
The parameters for the power-law component were fixed to the best-fit values of the result described above in the model-fitting. 
We note that the CXB and GH are affected by the Galactic hydrogen absorption while the LHB is not.
In our observed regions, the Galactic absorption column density was $\mathrm{n_H} = 7.58\times10^{20}\,\si{cm}^{-2}$ according to \verb|nh| in the HEASoft package, which we adopted.
We adopted the solar abundance table given by \citet{Anders1989}; the abundances of carbon, nitrogen, neon and iron were set to be free while those of the other elements were fixed to unity in our model fitting.
All the abundance parameters were linked between the two thermal plasma models.
Figure \ref{fig:stable_spec} shows the spectra and resultant best-fit models and table \ref{table:stable_param} summarizes the fitting results. 
The fitting was satisfactory with a reduced chi-square of 1.06.

\begin{table}[H]
  \tbl{Results of the model fitting to the \textit{stable}-period spectra.}{%
  \begin{tabular}{cccc}
      \hline \hline 
                          Model                         &                           Parameters                              &         &                                               \\  \hline
    CXB (power-law)                              & $N_\mathrm{H} \mathrm{[10^{22}~\cm^{-2}]}$ &          &  0.0758 (frozen)                    \\
                                                             & Photon Index $\Gamma$                                  &          &  1.41 (frozen)                        \\
                                                             & Normalization\footnotemark[$*$]                       &          &  10.1 (frozen)                        \\ \hline
    GH (vapec)                                      & $N_\mathrm{H} \mathrm{[10^{22}~\cm^{-2}]}$  &          &  0.0758 (frozen)                    \\
                                                             & $kT$ [keV]                                                         &          &  $0.297^{+0.013}_{-0.012}$ \\
                                                             & abundance\footnotemark[$\dag$]                     &    C    &  $1.26^{+0.76}_{-0.51}$       \\
                                                             &                                                                           &    N    &  $2.70^{+0.43}_{-0.40}$       \\                                   
                                                             &                                                                           &    O    &  1.00 (frozen)                       \\
                                                             &                                                                           &   Ne   &  $1.26^{+0.20}_{-0.18}$       \\
                                                             &                                                                           &   Fe   &  $0.45^{+0.10}_{-0.08}$        \\
                                                             & Normalization\footnotemark[$\ddag$]               &           &  $32.2^{+3.5}_{-6.5}$           \\ \hline
   LHB (vapec)\footnotemark[$\S$]       & $kT$ [keV]                                                        &           &  $0.130^{+0.016}_{-0.010}$  \\
                                                             & Normalization\footnotemark[$\ddag$]               &           &  $45.9^{+12.2}_{-8.9}$          \\ \hline \hline
                                                             & $\chi^2$/d.o.f                                                     &           &  325.02 / 308                         \\ \hline \hline
    \end{tabular}}
\label{table:stable_param}
\begin{tabnote}
\footnotemark[$*$] In units of $\mathrm{photons\,s^{-1}\,cm^{-2}\,keV^{-1}\,str^{-1}}$ at $1\keV$. \\ 
\footnotemark[$\dag$] These values are abundances with respect to solar. \\
\footnotemark[$\ddag$] In units of $(4\pi)^{-1}~D_{\mathrm{A}}^{-2}~(1+z)^{-2}~10^{-14} \int n_{\mathrm{e}}~ n_{\mathrm{H}}~dV$ per steradian, 
where $D_{\mathrm{A}}$ is the angular size distance to the source (cm), and $n_{\mathrm{e}}$ and $n_{\mathrm{H}}$ are the electron and hydrogen densities ($\mathrm{cm^{-3}}$), respectively.  \\
\footnotemark[$\S$] Each abundance parameter is linked to that of the GH model.  \\
\end{tabnote}
\end{table}

\subsubsection{\textit{active} period}

\begin{figure}
 \begin{center}
  \includegraphics[width=13cm]{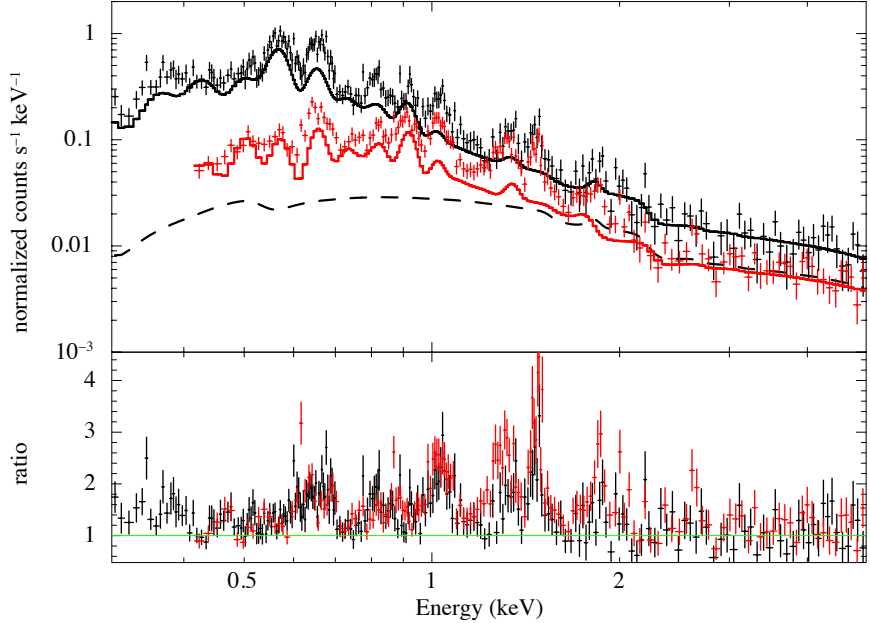}
 \end{center}
\caption{ Spectra of the (black) BI CCD and (red) FI CCDs in the \textit{active} period. The solid lines show the model consisting of the 2006 model and, for the BI-CCD spectrum only, an unabsorbed power-law (dashed line). }
\label{fig:active_spec_fit_NoGauss}
\end{figure}

We then analyzed the spectra in the \textit{active} period to evaluate quantitatively the difference from the spectra in the \textit{stable} period
 (hereafter, the spectra and best-fit model of the latter are referred to as the 2006 spectra and model, respectively).
Since these event data were extracted from an identical celestial region, we can get rid of the ambiguity due to the difference of the observational directions in the GSE coordinates.
Following the method for the model fitting of the 2006 spectra, we first fitted the spectra of the \textit{active} period in the $2.0$--$5.0\keV$~only with a power-law of which the parameters were fixed to the same as those of the 2006 model. 
We found that whereas the best-fit parameters were consistent with those of the 2006 model  with the FI-CCD spectrum, those of the BI-CCD spectrum showed an excess over the 2006 model.
A similar excess in the BI-CCD spectrum was also reported in the past SWCX studies with Suzaku (e.g., \cite{Fujimoto2007}; \cite{Ezoe2011}).
Since the excess is only seen in the BI spectrum, it is probably due to particle backgrounds such as soft protons, as discussed in \citet{Ezoe2011}.
Regarding spectral fitting, \citet{Fujimoto2007} dealt with this feature by adding an unabsorbed power-law to the blank-sky model.
Accordingly, we added an unabsorbed power-law to the 2006 model in order to address the discrepancy.
Figure \ref{fig:active_spec_fit_NoGauss} shows the BI- and FI-CCD spectra and the 2006 model plus the additional power-law.
Many Gaussian-like residuals, mainly below $2\keV$, are still clearly visible, after including the additional power-law component.

\begin{figure}
 \begin{center}
  \includegraphics[width=13cm]{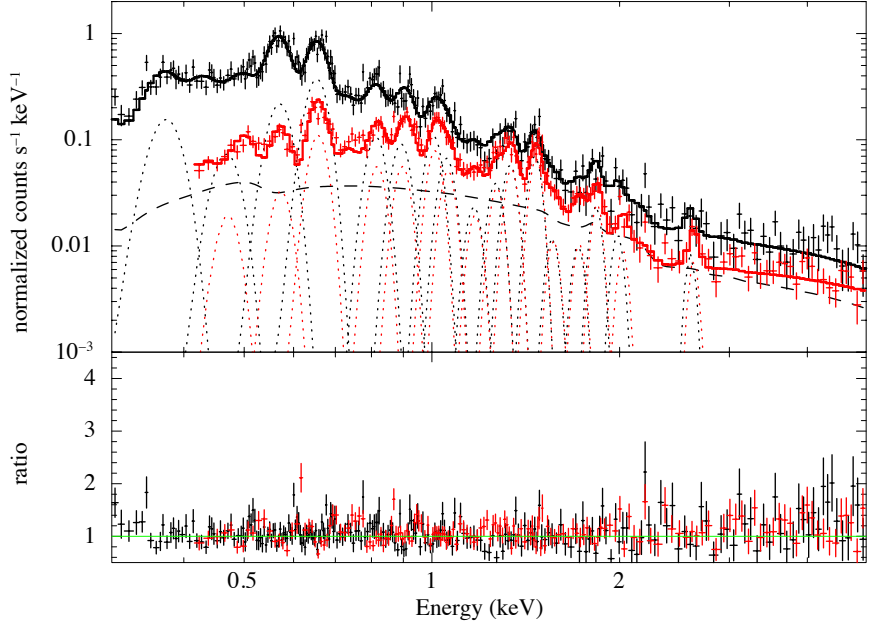}
 \end{center}
\caption{Same as figure \ref{fig:active_spec_fit_NoGauss}, but the model contains additional 17 Gaussians (dotted lines). The slight offset between the Gaussians for the BI- and FI-CCD spectra are due to the difference of energy gain.}
\label{fig:active_spec_fit_Gauss}
\end{figure}

Then, we added Gaussian models one by one to the spectral model to eliminate the significant residuals and repeated model-fitting.
In the fittings, the center energy and normalization of each Gaussian were allowed to vary but were set to be common (\textit{i.e.}, linked) between the BI- and FI-CCD spectra.
Consequently, we obtained an acceptable fit by adding 17 Gaussian models.
Figure \ref{fig:active_spec_fit_Gauss} shows the best-fit model with individual components.
Residuals that are still seen in figure  \ref{fig:active_spec_fit_Gauss} may be due to systematic uncertainties of the line profiles.
The parameters of the additional power-law and Gaussians are summarized in tables \ref{table:active_spec_fit} and \ref{table:gauss}.
It should be noted that the additional power-law component is response-folded in our analysis. On the other hand, the soft proton contamination in the XMM-Newton observations can be approximated by a broken power-law model which is not folded through the detector response (\cite{Kuntz2008}).
We also confirm that our results shown in table \ref{table:gauss} remain almost unchanged (even for the C lines the decrease of the normalization is less than 30\%), even if we substitute the response-unfolded power-law for the response-folded one.
The center energy of many of the Gaussians was found to be consistent with the energies of the characteristic X-rays from H-like or He-like ions of one of the relatively abundant elements.
Notably, we identified the $459\eV$~line (C \emissiontype{VI} 4p to 1s), which is one of the main features of the SWCX (\cite{Fujimoto2007}).
Therefore we concluded that the residuals in the fitting with the 2006 model were attributed to the SWCX emission.

\begin{table}
  \tbl{Results of the  model fitting to the \textit{active}-period spectra.}{%
  \begin{tabular}{ccc}
      \hline \hline
                          Model                            &  Parameters                                                    &                                        \\  \hline
                     2006 model                        &  Described in table \ref{table:stable_param}.  &                                        \\  \hline
              Additional power-law                 &  Photon Index $\Gamma$                               & $1.47^{+0.20}_{-0.21}$  \\ 
                                                                &  Normalization\footnotemark[$*$]                   & $8.26^{+1.29}_{-1.00}$  \\ \hline
           Additional 17 Gaussians              &  Described in table \ref{table:gauss}.              &                                        \\  \hline \hline
                                                                & $\chi^2$/d.o.f                                                  &       524.60 / 369             \\ \hline \hline
    \end{tabular}}\label{table:active_spec_fit}
\begin{tabnote}
\footnotemark[$*$] In units of $\mathrm{photons\,cm^{-2}\,s^{-1}\,keV^{-1}\,str^{-1}}$ at $1\keV$. \\
\end{tabnote}
\end{table}

\begin{table}
  \tbl{Best-fit parameters of the Gaussians and identification of the lines.}{%
  \begin{tabular}{ccccc}
      \hline \hline
   Gaussian\footnotemark[$*$] &
           Center Energy [eV]     &  Normalization\footnotemark[$\dag$]  & $f_{\mathrm{X}}$\footnotemark[$\ddag$] &  Line Identification \\ \hline
1    &    $374^{+7}_{-10}$       &             $11.6^{+2.8}_{-3.1}$               &           $7.42^{+1.79}_{1.96}$                   &  C \emissiontype{VI} 2p to 1s (368\,eV) \\
2    &    $468^{+10}_{-10}$     &	          $2.73^{+1.00}_{-1.00}$             &           $2.18^{+0.80}_{-0.80}$                  &  C \emissiontype{VI} 4p to 1s (459\,eV) \\ 
3    &    $571^{+6}_{-5}$         &           $7.81^{+1.58}_{-1.48}$             &           $7.59^{+1.54}_{-1.44}$                  &  O \emissiontype{VII} \\ 
4    &    $656^{+2}_{-6}$         &           $9.86^{+0.86}_{-0.95}$             &             $11.0^{+1.0}_{-1.1}$                    &  O \emissiontype{VIII} 2p to 1s (653\,eV)  \\ 
5    &    $814^{+7}_{-6}$         &           $2.20^{+0.38}_{-0.36}$             &           $3.05^{+0.53}_{-0.50}$                  &  O \emissiontype{VIII} or Fe \emissiontype{XVII} \\ 
6    &    $895^{+8}_{-10}$       &           $1.78^{+0.32}_{-0.33}$             &           $2.71^{+0.48}_{-0.50}$                  &  Ne \emissiontype{IX} or Fe \emissiontype{XVIII} \\ 
7    &    $1008^{+11}_{-4}$	     &           $2.01^{+0.30}_{-1.02}$             &           $3.45^{+0.51}_{-1.76}$                  &  Ne \emissiontype{X} 2p to 1s (1022\,eV)  \\ 
8    &    $1060^{+18}_{-8}$     &           $1.21^{+0.62}_{-0.68}$             &           $2.19^{+1.11}_{-1.23}$                  &  Ne \emissiontype{IX} ? \\
9    &    $1172^{+18}_{-19}$   &           $0.42^{+0.13}_{-0.14}$             &           $0.83^{+0.27}_{-0.27}$                  &    -   \\
10  &    $1270^{+10}_{-14}$   &           $0.72^{+0.15}_{-0.14}$             &           $1.55^{+0.32}_{-0.30}$                  &   -    \\
11  &    $1339^{+7}_{-7}$	     &           $1.04^{+0.16}_{-0.16}$             &           $2.37^{+0.36}_{-0.37}$                  &  Mg \emissiontype{XI} \\
12  &    $1468^{+5}_{-5}$       &           $1.26^{+0.15}_{-0.14}$             &           $3.15^{+0.37}_{-0.36}$                  &  Mg \emissiontype{XII} 2p to 1s (1473\,eV) \\ 
13  &    $1564^{+20}_{-17}$   &           $0.28^{+0.12}_{-0.12}$             &           $0.75^{+0.33}_{-0.32}$                  &  Mg \emissiontype{XI} \\
14  &    $1722^{+22}_{-25}$   &           $0.23^{+0.12}_{-0.10}$             &           $0.69^{+0.35}_{-0.29}$                  &  Mg \emissiontype{XII} 3p to 1s (1745\,eV) \\ 
15  &    $1836^{+3}_{-4}$       &           $0.55^{+0.09}_{-0.12}$             &           $1.72^{+0.29}_{-0.38}$                  &  Si \emissiontype{XIII}  \\
16  &    $2006^{+27}_{-27}$   &           $0.31^{+0.11}_{-0.11}$              &           $1.04^{+0.39}_{-0.39}$                  &  Si \emissiontype{XIV} 2p to 1s (2006\,eV) \\ 
17  &    $2616^{+23}_{-23}$   &           $0.33^{+0.12}_{-0.12}$             &            $1.46^{+0.54}_{-0.53}$                 &  S \emissiontype{XVI} 2p to 1s (2623\,eV) \\ \hline \hline
    \end{tabular}}\label{table:gauss}
\begin{tabnote}
\footnotemark[$*$] The width of the lines are fixed to 0. \\
\footnotemark[$\dag$] In units of $\mathrm{photons\,cm^{-2}\,s^{-1}\,str^{-1}}$.  \\ 
\footnotemark[$\ddag$] $f_{\mathrm{X}}$ is energy flux in units of $10^{-13}\,\mathrm{erg\,s^{-1}\,cm^{-2}}$.\\ 
\end{tabnote}
\end{table}

\section{Discussion}

\subsection{Significance of the sulfur line}

In \S3, we have evaluated the SWCX lines from several ions with 17 Gaussian models.
Notably, we have found a line at around $2.62\keV$, the energy of which coincides with the S \emissiontype{XVI} Ly$\alpha$ line.
No previous studies have reported a positive detection of a S \emissiontype{XVI} line from the SWCX emission.
Our detection is a significant discovery if it is genuinely a S \emissiontype{XVI} line. Here we discuss how plausible it is the case.\newline\indent

The SWCX emission is known to contain not only He$\alpha$ and Ly$\alpha$ lines but also various lines caused by other transitions.
In case of the $2.62\keV$ line, there is a possibility that this line is attributed to Si \emissiontype{XIV} Ly$\zeta$ and Ly$\eta$, of which the energies are at around $2.62\keV$.
In order to evaluate the contribution of the characteristic cascades, we incorporate a set of the AtomDB Charge Exchange (ACX) models (\cite{Smith2014}) \footnote{http://www.atomdb.org/CX/}
 instead of a set of simple Gaussians, fit the spectra with the model, and see how the residuals in figure~\ref{fig:active_spec_fit_NoGauss} are explained, as follows. 
The residuals in figure~\ref{fig:active_spec_fit_NoGauss} can be mostly attributed to the characteristic lines from seven elements of carbon, oxygen, neon, magnesium, silicon, sulfur, and iron (table \ref{table:gauss}).
Accordingly, our new model consists of, in addition to the continuum of the 2006 model and an additional power-law, 7 \verb|vacx| models (the variable abundance version of the ACX model), each of which represents the charge exchange emission from a separate element (\textit{n.b.}, we set the other abundances to a fixed value of 0).
Since we have no way to measure the absolute hydrogen abundance in the plasma (the \verb|vacx| model has no continuum), only the relative abundances can be obtained in principle.
Therefore, we link the normalizations of all the \verb|vacx| components and fix the oxygen abundance to unity in the fitting.
We perform model-fitting of the spectra in the \textit{active} period and obtain the relative abundances of the six elements to oxygen and ion population $T_{\mathrm{z}}$,
 which indicates the ion population as though it were in collisional ionization equilibrium created by electrons at temperature $T_\mathrm{z}$.
These ion population derived by SWCX do not necessarily coincide with the real ion population; our observation in the X-ray band is only sensitive to highly charged ions.

\begin{figure}
 \begin{center}
  \includegraphics[width=13cm]{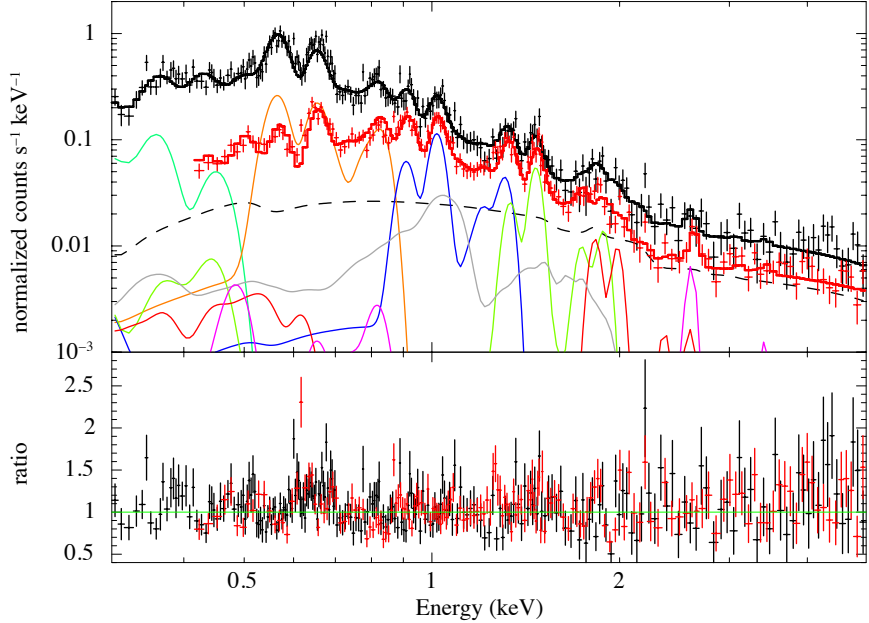}
 \end{center}
\caption{Same as figure \ref{fig:active_spec_fit_Gauss}, but the model has 7 ACX models instead of Gaussians (only those for the BI spectrum are depicted). The line color distinguishes the element of each ACX model: (green)\ C, (orange) O,  (gray) Fe, (blue) Ne, (lime-green) Mg, (red) Si, and (magenta) S.}
\label{fig:active_spec_fit_vacx}
\end{figure}

Table~\ref{table:active_spec_fit_vacx} lists the fitting results and Figure \ref{fig:active_spec_fit_vacx} shows the best-fit model. 
The relative abundances of the elements to oxygen are found to be close to the solar abundances, except for that of carbon, which is significantly higher (carbon-rich) than the solar abundance. 
We conjecture that the apparent high-carbon abundance is probably due to an effect of the contamination from the heliospheric SWCX emission.
We also note that the tendency of the low abundances of silicon and sulfur are consistent with an other observational result of the same solar flare with Suzaku, using Earth X-ray albedo (\cite{Katsuda2020}).
This result, where the X-rays from the Si \emissiontype{XIV} line are fully taken into account, implies that the ACX model of sulfur is needed to explain the observed spectra at a confidence level of more than 3$\sigma$.
Hence we conclude that the detected $2.62\keV$ line is S \emissiontype{XVI} Ly$\alpha$, which is the first detection of the line from the SWCX emission.

\begin{table}  
\tbl{Best-fit parameters with the 2006 model plus additional power-law and seven ACX models.}{
\begin{tabular}{cccc}
   \hline \hline
   \multicolumn{2}{c}{Model}       &  \multicolumn{2}{c}{Parameters}                                                                     \\ \hline \hline                     
         2006 model           &          &  \multicolumn{2}{c}{Described in table \ref{table:stable_param}.}                   \\ \hline
                                       &          &  Photon Index $\Gamma$                 &   Normalization\footnotemark[$*$]          \\ \hline
  Additional power-law    &          &      $1.21_{-0.36}^{+0.35}$                  &           $6.33_{-1.72}^{+1.74}$                  \\ \hline
                                       &          &   abundance \footnotemark[$\dag$]  &  Ion population $T_\mathrm{z}$ [keV]    \\  \hline
               vacx                 &  C     &                      $> 4.89$                       &        $0.075_{-0.005}^{+0.008}$             \\
                                       &  O     &                      1 (fixed)                        &        $0.210_{-0.009}^{+0.010}$              \\
                                       &  Ne   &          $1.22_{-0.24}^{+0.28}$            &        $0.526_{-0.033}^{+0.046}$             \\
                                       &  Mg   &          $1.67_{-0.33}^{+0.41}$            &         $0.988_{-0.098}^{+0.146}$            \\
                                       &  Si     &       $0.729_{-0.300}^{+0.371}$         &            $1.18_{-0.20}^{+0.39}$               \\
                                       &  S      &       $0.619_{-0.320}^{+0.362}$         &                      $> 2.35$                            \\
                                       &  Fe    &          $1.39_{-0.49}^{+0.61}$            &         $1.16_{-0.07}^{+0.12}$                  \\   \hline \hline     
                                       &          &              $\chi^2$/d.o.f                        &                  595.56 / 389                         \\  \hline \hline   
   \end{tabular}}\label{table:active_spec_fit_vacx}
\begin{tabnote}
\footnotemark[$*$] In units of $\mathrm{photons\,cm^{-2}\,s^{-1}\,keV^{-1}\,str^{-1}}$ at $1\keV$. \\ 
\footnotemark[$\dag$] These values are abundances with respect to solar. \\
\end{tabnote}
\end{table}

\subsection{Correlation with the CME}

\begin{figure}
 \begin{center}
  \includegraphics[width=13cm]{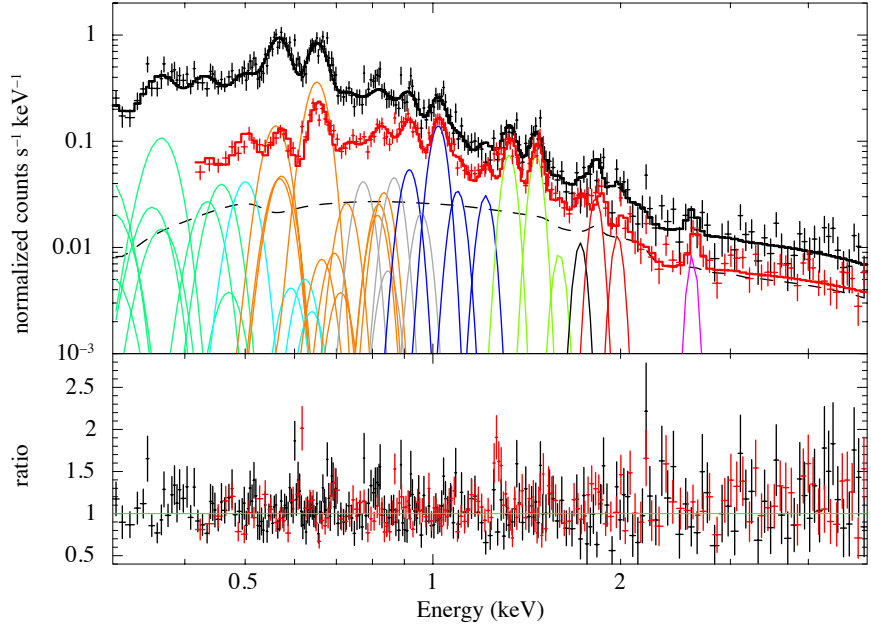}
 \end{center}
\caption{Same as figure \ref{fig:active_spec_fit_NoGauss}, but the model additionally contains the same SWCX model as in \citet{Carter2010} and the S \emissiontype{XVI} line (only those for the BI spectrum are depicted). The center energy of each Gaussian is fixed to the transition energy of each ion. The line color shows the type of the element species: (green) C, (cyan) N, (orange) O, (gray) Fe, (blue) Ne, (lime-green) Mg, (black) Al, (red) Si, and (magenta) S. }
\label{fig:active_spec_fit_FixedGauss}
\end{figure}

We have shown that Suzaku detected various kinds of the SWCX lines during the \textit{active} period.
Given that indicators of an ICME were observed by in-situ measurements at the same time, this SWCX emission is likely to be associated with the ICME.
Some past similar studies examined the correlation between the observed X-ray light curves and arrival of an ICME to establish their association (e.g. \cite{Ezoe2010, Ishi2019}).
However, since the observed X-rays during the observation in the \textit{active} period did not show much time fluctuation, the method would yield nothing significant in our case.
Instead, we evaluate the plasma metal-abundance in the same way as in a previous study of spectrally-rich ICME-driven SWCX emission (\cite{Carter2010}).
They took account of 33 emission lines from carbon, nitrogen, and oxygen, of which the emission cross sections are listed in table 2 in \citet{Bodewits2007}.
The normalization of the line with the largest cross section in each ion species was set as a free parameter in the fitting, while those of the other lines were determined according to the relative emission cross-sections.
In our case, we adopt the cross section at the solar wind speed of 1000~$\mathrm{km\,s^{-1}}$ which is close to the average speed during our observation period (915\,$\mathrm{km\,s^{-1}}$).
As for the other line parameters, we add Gaussians for the ions at the fixed transition energies which were detected in \citet{Carter2010} and for the S \emissiontype{XVI} at $2.62\keV$.
Then we fit the spectra in the \textit{active} period with the model. 
Table~\ref{table:fixedgauss} summarizes the fitting result and figure \ref{fig:active_spec_fit_FixedGauss} shows the best-fit spectra.
We find that this model also can explain the observed spectra.

\begin{figure}
 \begin{center}
  \includegraphics[width=15cm]{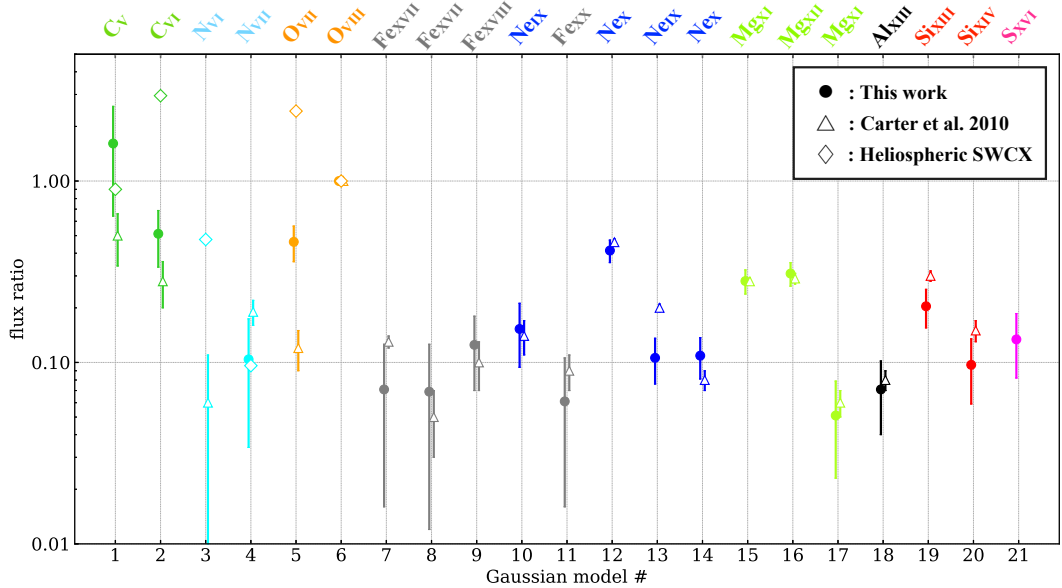}
 \end{center}
\caption{Energy flux ratio of the line to the O \emissiontype{VIII} line ($654\eV$). Circles and triangles indicate our results and those by \citet{Carter2010}, respectively. Diamonds indicate the heliospheric SWCX case calculated with the slow solar wind abundance given by \citet{Schwadron2000}.}
\label{fig:ratio_toO}
\end{figure}

From the fitting result, we calculate the energy flux of each ion species and plot the energy flux ratios of the lines to the $\mathrm{O_{VIII}}$ line ($654\eV$) in figure \ref{fig:ratio_toO}.
In the figure, we also plot for reference the results by \citet{Carter2010} and the calculated energy flux ratio for the heliospheric SWCX, where the same cross sections and slow equatorial solar-wind abundance as given by \citet{Schwadron2000} are assumed.
Our result is found to roughly agree with that of the ICME-driven case \citep{Carter2010} except for carbon, nitrogen, and oxygen.
We conjecture that the discrepancies with regard to carbon, nitrogen, and oxygen are due to a potential increase of the heliospheric SWCX component, 
given that the intensities of the heliospheric SWCX can be different between the \textit{stable} and \textit{active} periods.
The detection of the lines from Fe \emissiontype{XVII}, Fe \emissiontype{XVIII}, and Fe \emissiontype{XX} indicates the existence of highly-ionized iron;
it is consistent with the characteristic of the iron state in ICMEs (summarized in \cite{Zurbuchen2006}), whereas the stable slow solar wind barely has such ions.
These features also support that this event is ICME-driven.

The first detection of the emission line from S \emissiontype{XVI} suggests this to be the most spectrally-rich SWCX event to date ever reported 
(\textit{n.b.}, there is a possibility that the energy range above $2\keV$ was not fully investigated in some of the SWCX studies and that line features were overlooked).
The flare which triggered the ICME associated with the event is of the X6 class, which is one of the most intense flares in the ICME-driven SWCX samples.
This fact may suggest that the solar flare magnitude is related to the ionization states in the ICME, as pointed out by \citet{Reinard2005}.

\begin{table}
  \tbl{Best-fit parameters in model fitting with the fixed-center-energy Gaussians.}{%
  \begin{tabular}{cccc}
      \hline \hline
   Gaussian\footnotemark[$*$] &    Center Energy [eV]    & Normalization\footnotemark[$\dag$] &      Line Identification      \\ \hline
                                           1    &               299                 &        $33.6^{+20.0}_{-20.0}$             &               C \emissiontype{V}        \\
                                           2    &               367                 &        $8.71^{+2.87}_{-2.89}$             &               C \emissiontype{VI}       \\
                                           3    &               420                 &                        -                                 &               N \emissiontype{VI}       \\ 
                                           4    &                500                &        $1.30^{+0.86}_{-0.87}$              &               N \emissiontype{VII}      \\  
                                           5    &               561                 &        $5.15^{+1.03}_{-1.04}$              &               O \emissiontype{VII}      \\ 
                                           6    &               653                 &        $9.57^{+0.93}_{-0.93}$              &               O \emissiontype{VIII}     \\ 
                                           7    &               730                 &        $0.60^{+0.47}_{-0.47}$              &               Fe \emissiontype{XVII}  \\ 
                                           8    &               820                 &        $0.52^{+0.43}_{-0.43}$              &               Fe \emissiontype{XVII}  \\ 
                                           9    &               870                 &        $0.90^{+0.38}_{-0.38}$              &               Fe \emissiontype{XVIII} \\ 
                                           10  &               920                 &        $1.04^{+0.39}_{-0.39}$              &               Ne \emissiontype{IX}     \\
                                           11  &               960                 &        $0.39^{+0.29}_{-0.29}$              &               Fe \emissiontype{XX}    \\
                                           12  &               1022               &        $2.53^{+0.26}_{-0.26}$              &               Ne \emissiontype{X}      \\ 
                                           13  &               1100               &        $0.60^{+0.16}_{-0.16}$              &              Ne \emissiontype{IX}      \\
                                           14  &               1220               &        $0.56^{+0.14}_{-0.14}$              &               Ne  \emissiontype{X}     \\
                                           15  &               1330               &        $1.32^{+0.15}_{-0.15}$              &               Mg \emissiontype{XI}     \\
                                           16  &               1470               &        $1.31^{+0.15}_{-0.15}$              &               Mg \emissiontype{XII}    \\
                                           17  &               1600               &        $0.20^{+0.11}_{-0.11}	$              &                Mg \emissiontype{XI}     \\
                                           18  &               1730               &        $0.26^{+0.11}_{-0.11}$              &                Al \emissiontype{XIII}     \\ 
                                           19  &               1850               &        $0.69^{+0.15}_{-0.15}$              &                Si \emissiontype{XIII}    \\
                                           20  &               2000               &        $0.30^{+0.12}_{-0.12}$              &               Si \emissiontype{XIV}     \\
                                           21  &               2623               &        $0.32^{+0.12}_{-0.12}$              &                S \emissiontype{XVI}     \\ \hline
    \end{tabular}}\label{table:fixedgauss}
\begin{tabnote}
\footnotemark[$*$] The width of the lines are fixed to 0. \\
\footnotemark[$\dag$] In units of $\mathrm{photons\,cm^{-2}\,s^{-1}\,str^{-1}}$.  \\ 
\end{tabnote}
\end{table}

\subsection{SWCX as a new tool to detect highly-ionized ions in the solar wind}

From the eruption at the low corona to propagation to interplanetary space, a CME evolves and its properties change.
In particular, the CME plasma in the early phase of the evolution (up to a few $R_{\odot}$) undergoes a continuous heating process; 
signs of heating have been observed by Ultraviolet Coronagraph Spectrometer (UVCS) onboard SOHO, which provides valuable information to constrain the heating model (e.g. \cite{Akmal2001}, \cite{Landi2010}).
Even though the evolution of CMEs have been studied extensively with observations and simulations for a few decades, the heating mechanism still remains an open question.

The ionic charge-state distribution in the ICME is also an effective tool to diagnose its plasma state.
As CME ejecta are adiabatically expanding toward the outer region, the density of the ejecta gradually decreases.
Once the density has dropped sufficiently low, the ejecta become collision-less and then the ionization state of each ion is fixed at the specific distance from which no ionization or recombination takes place
 (so-called  ``freeze-in height'') while it is transferred away to the heliosphere.
Hence, the ionization state of ions in an ICME preserves its post-heating plasma information at the freeze-in height.
It implies that in-situ measurements of these ions can also constrain the evolution history of CMEs, as discussed in e.g., \citet{Gruesbeck2011} and \citet{Rivera2019}.

\begin{figure}
 \begin{center}
  \includegraphics[width=15cm]{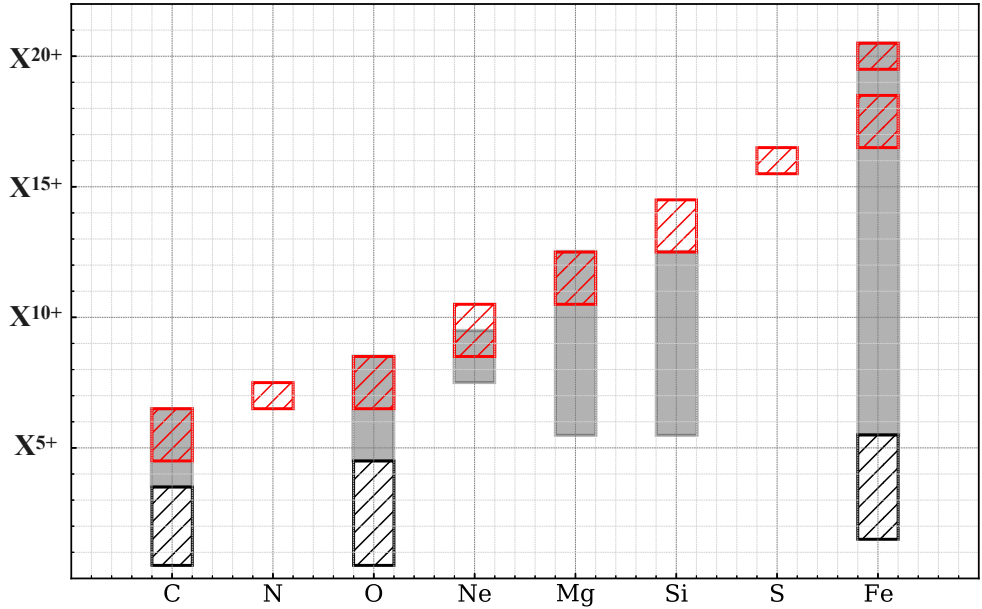}
 \end{center}
\caption{Detectable charge states of C, N, O, Ne, Mg, Si, S, and Fe  with the currently available observations. 
Gray shaded areas show the available charge state distributions in the SWICS standard dataset. Red and black hatched areas show the ranges obtained in our SWCX observation 
 and \citet{Gilbert2012}, respectively.}
\label{fig:detectable_ions}
\end{figure}

The ionic charge-state distribution in ICMEs have been repeatedly observed with the Solar Wind Ion Composition Spectrometer (SWICS) onboard Advanced Composition Explorer (\cite{Gloeckler1998}).
The SWICS can obtain information of the plasma composition using three observable parameters for each ion count (energy per charge, time of flight, and residual energy).
The SWICS data are publicly distributed and serve as the standard dataset in this field, which contain information of several ions (gray regions in figure \ref{fig:detectable_ions}). 
Although the SWICS is able to detect a wide range of energy, some ion species are deliberately excluded from the public standard dataset due to complexity of data analyses.
\citet{Gilbert2012} reported that they detected low-charged ions in ICMEs with a new analysis method, extending the observed parameter space of the ion species in ICMEs (black hatched regions in figure \ref{fig:detectable_ions}).
However, it is still difficult for the SWICS to observe some H-like or He-like ions due to poor statistics.

The SWCX emission from a variety of highly charged ions provides information of plasma composition (see \S1).
In particular, some ICME-driven SWCX events have emission lines from ions that are hardly detected by SWICS\ but can be detected with the data similar to those used in this work.
Although there is a downside in our method that the accurate cross section of the SWCX is required to obtain the absolute flux of each ion, 
we suggest that ICME-driven SWCXs can provide a unique opportunity to detect these highly-charged ions in ICMEs.
So far, in-situ measurements have been the only way to obtain the physical parameters useful to study the plasma state at the freeze-in point.
The data and method used in this work serve as a new and complementary way for it.

Though Suzaku and XMM-Newton have made great contributions to SWCX research, 
the energy resolutions of the spectroscopic imagers onboard these satellites are not sufficiently high to resolve the complex line structures, especially below $1\keV$.
Some of the planned future X-ray astronomical satellites are expected to address the issue greatly better. 
The X-Ray Imaging and Spectroscopy Mission (XRISM), scheduled to be launched in Japanese fiscal year 2022, has a non-dispersive soft X-ray spectrometer with a high spectral resolution 
(the full-width at half-maximum $<$ 7$\eV$ at $6\keV$) over the $0.3$--$12.0\keV$ band-pass (\cite{Tashiro2020}). Spectroscopy with XRISM can reveal the line structure for the first time.
Also, Advanced Telescope for High-ENergy Astrophysics (Athena) can also contribute to the study of ICMEs through observations of the SWCX if it is put at the L1 Lagrange point.

Finally, charge exchange process happens not only in our solar system but also in supernova remnants, starforming galaxies, and galaxy clusters.
Hence, the understanding of the physical process is important in a wide field of astrophysics, as discussed in \citet{XRISM2020}.

\section{Conclusions}

We analyzed the SN1006 background data observed with Suzaku on 2005 September 11, using another observation of the same field at a different epoch as the reference.
We found that the data contained an additional soft-X-ray component compared with the reference data, which were taken when the solar activity was stable.
With spectral model fitting, we revealed the additional component to be multiple emission lines, which can be explained as SWCX emission.
The observation period coincided with the passage of an ICME associated with a X6 flare that had happened shortly before the Suzaku observation.
The coincidence, as well as the fact the characteristics of the emission lines were consistent with those of the ICME-driven SWCX events reported in the past, strongly implies that this SWCX event was ICME-driven.
Notably, we detected an emission line from S \emissiontype{XVI} with a confidence level of more than 3$\sigma$ for the first time as one from a SWCX, 
the fact of which suggests that this is the most spectrally-rich SWCX event ever reported to date.
Note that some of the ions, including highly-ionized sulfur, that we detected cannot be detected by currently-available in-situ solar-wind monitoring due to poor statistics.
We propose that the SWCX is a new tool for diagnosing the ICME plasma, the understanding of which is important not only in studying X-ray diffuse background but also understanding the coronal heating mechanism.

\begin{ack}
The authors gratefully acknowledge the use of data obtained from the Suzaku and ACE satellites, NASA/GSFC's Space Physics Data Facility's OMNIWeb service, and SOHO LASCO CME catalog.
The CME catalog is generated and maintained at the CDAW Data Center by NASA and The Catholic University of America in cooperation with the Naval Research Laboratory.
SOHO is a project of international cooperation between ESA and NASA.
This work was supported by Japan Society for the Promotion of Science (JSPS) KAKENHI Grant Nos.\ 20J20685 (KA), 
20H00175 (HM), 18J20523 (TY), 19K21884, 20H01941, 20H01947, 20KK0071 (HN), 18K18767, 19H00696, 19H01908, 20H00176 (KH), 20K20935 (SK), 19J20910 (DI) and 20H00177 (YE).
This work was also partly supported by the Mitsubishi Foundation Research Grants in the Natural Sciences 201910033 (KH), 
Leading Initiative for Excellent Young Researchers, MEXT, Japan (SK), and Toray Science and Technology Grant (YE).
\end{ack}


\begin{thebibliography}{}

\bibitem[Akmal et al. (2001)]{Akmal2001} Akmal, A., Raymond, J. C., Vourlidas, A., Thompson, B., Ciaravella, A., Ko, Y. -K., Uzzo, M. and Wu, R. 2001, \apj, 553, 922
\bibitem[Anders \& Grevesse(1989)]{Anders1989} Anders, E. and Grevesse, N. 1989, Geochim. Cosmochim. Acta, 53, 197
\bibitem[Bodewits et al.(2007)]{Bodewits2007} Bodewits, D., et al. 2007, \aap, 469, 1183
\bibitem[Burlaga et al.(1981)]{Burlaga1981} Burlaga, L., Sittler, E., Mariani, F. and Schwenn, R. 1981, \jgr, 86, 6673
\bibitem[Carter \& Sembay(2008)]{Carter2008} Carter, J. A. and Sembay, S. 2008, \aap, 489, 837
\bibitem[Carter et al.(2010)]{Carter2010} Carter, J. A., Sembay, S. and Read, A. M. 2010, \mnras, 402, 867
\bibitem[Carter et al.(2011)]{Carter2011} Carter, J. A., Sembay, S. and Read, A. M. 2011, \aap, 527, A115
\bibitem[Chen(2011)]{Chen2011} Chen, P. F. 2011, Living Rev. \solphys, 8, 1
\bibitem[Colaninno \& Vourlidas(2009)]{Colaninno2009} Colaninno, Robin C. and Vourlidas, Angelos 2009, \apj, 698, 852
\bibitem[Cox(1998)]{Cox1998} Cox, D. P. 1998, in IAU Collq. 166, the local bubble and beyond, ed. Breitschwerdt, D., Freyberg, M. J., \& Truemper, J., Lecture Notes in Physics, 506 (Berlin: Springer Verlag), 121
\bibitem[Cravens(1997)]{Cravens1997} Cravens, T. E. 1997, Geophysical Research Letters, 24, 105
\bibitem[De Luca \& Molendi(2004)]{Luca2004} De Luca, A. and Molendi, S. 2004, \aap, 419, 837
\bibitem[Ezoe et al.(2010)]{Ezoe2010} Ezoe, Y., Ebisawa, K., Yamasaki, N. Y., Mitsuda, K., Yoshitake, H., Terada, N., Miyoshi, Y. and Fujimoto, R. 2010, \pasj, 62, 981
\bibitem[Ezoe et al.(2011)]{Ezoe2011} Ezoe, Y., Miyoshi, Y., Yoshitake, H., Mitsuda, K., Terada, N., Oishi, S. and Ohashi, T. 2011, \pasj, 63, S691
\bibitem[Freyberg(1998)]{Freyberg1998} Freyberg, M. J. 1998, in IAU Collq. 166, the local bubble and beyond, ed. Breitschwerdt, D., Freyberg, M. J., \& Truemper, J., Lecture Notes in Physics, 506 (Berlin: Springer Verlag), 113
\bibitem[Fujimoto et al.(2007)]{Fujimoto2007} Fujimoto, R., et al. 2007, \pasj, 59, S133
\bibitem[Gilbert et al. (2012)]{Gilbert2012} Gilbert, J. A., Lepri, S. T., Landi, E. and Zurbuchen, T. H. 2012, \apj, 751, 20
\bibitem[Gloeckler et al.(1998)]{Gloeckler1998} Gloeckler, G., et al. 1998, \ssr, 86, 497
\bibitem[Gopalswamy et al.(2009)]{Gopalswamy2009} Gopalswamy, N., Yashiro, S., Michalek, G., Stenborg, G., Vourlidas, A., Freeland, S. and Howard, R. 2009, Earth Moon and Planets, 104, 295
\bibitem[Gruesbeck et al.(2011)]{Gruesbeck2011} Gruesbeck, J. R., Lepri, S. T., Zurbuchen, T. H. and Antiochos, S. K. 2011, \apj, 730, 103
\bibitem[Ishi et al.(2017)]{Ishi2017} Ishi, D., Ishikawa, K., Ezoe, Y., Ohashi, T., Miyoshi, Y. and Terada, N. 2017, in The X-ray Universe 2017, ed. Ness, J-U, Migliari, S. 103
\bibitem[Ishi et al.(2019)]{Ishi2019} Ishi, D., Ishikawa, K., Numazawa, M., Miyoshi, Y., Terada, N., Mitsuda, K., Ohashi, T. and Ezoe, Y. 2019, \pasj, 71, 23
\bibitem[Ishisaki et al.(2007)]{Ishisaki2007} Ishisaki, Y., et al. 2007, \pasj, 59, S113
\bibitem[Katsuda et al.(2020)]{Katsuda2020} Katsuda, S., et al. 2020, \apj, 891, 126
\bibitem[Koyama et al.(2007)]{Koyama2007} Koyama, K., et al. 2007, \pasj, 59, S23
\bibitem[Kuntz \& Snowden(2008)]{Kuntz2008} Kuntz, K. D. and Snowden, S. L. 2008, \aap, 478, 575
\bibitem[Kuntz(2019)]{Kuntz2019} Kuntz, K. D. 2019, \aapr, 27, 1
\bibitem[Kushino et al.(2002)]{Kushino2002} Kushino, A., Ishisaki, Y., Morita, U., Yamasaki, N. Y., Ishida, M., Ohashi, T. and Ueda, Y. 2002, \pasj, 54, 327
\bibitem[Landi et al. (2010)]{Landi2010} Landi, E., Raymond, J. C., Miralles, M. P. and Hara, H. 2010, \apj, 711, 75
\bibitem[Lisse et al.(1996)]{Lisse1996} Lisse, C. M., et al. 1996, Science, 274, 205
\bibitem[Lopez(1987)]{Lopez1987} Lopez, R. E. 1987, \jgr, 92, 11189
\bibitem[Merka et al.(2005)]{Merka2005} Merka, J., Szabo, A., Slavin, J. A. and Peredo, M. 2005, \jgr, 110, A04202
\bibitem[Mitsuda et al.(2007)]{Mitsuda2007} Mitsuda, K., et al. 2007, \pasj, 59, S1
\bibitem[Reinard (2005)]{Reinard2005} Reinard, A. 2005, \apj, 620, 501
\bibitem[Rivera et al.(2019)]{Rivera2019} Rivera, Y. J., Landi, E., Lepri, S. T. and Gilbert, J. A. 2019, \apj, 874, 164
\bibitem[Schwadron \& Cravens(2000)]{Schwadron2000} Schwadron, N.A. and Cravens, T. E. 2000, \apj, 544, 558
\bibitem[Serlemitsos et al.(2007)]{Serlrmitsos2007} Serlemitsos, P. J., et al. 2007, \pasj, 59, S9
\bibitem[Shue et al.(1998)]{Shue1998} Shue, J. H., et al. 1998, \jgr, 103, 17691
\bibitem[Smith et al.(2014)]{Smith2014} Smith, R. K., Foster, A. R., Edgar, R. J. and Brickhouse, N. S. 2014, \apj, 787, 77
\bibitem[Snowden et al.(1994)]{Snowden1994} Snowden, S. L., McCammon, D., Burrows, D. N., and Mendenhall, J. A. 1994, \apj, 424, 714
\bibitem[Snowden et al.(2004)]{Snowden2004} Snowden, S. L., Collier, M. R. and Kuntz, K. D. 2004, \apj, 610, 1182
\bibitem[Takahashi et al.(2007)]{Takahashi2007} Takahashi, T., et al. 2007, \pasj, 59, S35
\bibitem[Tashiro et al. (2020)]{Tashiro2020} Tashiro, M., et al. 2020, \procspie, 11444, 293
\bibitem[Tawa et al.(2008)]{Tawa2008} Tawa, N., et al. 2008, \pasj, 60, S11
\bibitem[Tousey(1973)]{Tousey1973} Tousey, R. 1973, Space Res. XIII, 2, 713
\bibitem[Wang et al.(2006)]{Wang2006} Wang, Y., Xue, X., Shen, C., Ye, P., Wang, S. and Zhang, J. 2006, \apj, 646, 625
\bibitem[Wargelin et al.(2004)]{Wargelin2004} Wargelin, B. J., Markevitch, M., Juda, M., Kharchenko, V., Edgar, R. and Dalgarno, A. 2004, \apj, 607, 596
\bibitem[Webb \& Howard(1994)]{Webb1994} Webb, D. F. and Howard, R. A. 1994, \jgr, 99, 4201
\bibitem[Webb \& Howard(2012)]{Webb2012} Webb, D. F. and Howard, T. A. 2012, Living Rev. \solphys, 9, 3
\bibitem[XRISM Science Team(2020)]{XRISM2020} XRISM Science Team. 2020, arXiv e-prints, arXiv:2003.04962
\bibitem[Zurbuchen \& Richardson(2006)]{Zurbuchen2006} Zurbuchen, T. H. and Richardson, I. G. 2006, \ssr, 123, 31







\end{thebibliography}
\end{document}